\documentclass[preprint]{revtex4}
\usepackage{graphicx}
\usepackage{bm}

\def\bq{\begin{equation}}
\def\eq{\end{equation}}

\def\veps{\varepsilon}
\def\vphi{\varphi}

\def\cs{\cos\theta}

\def\E{{\bf{E}}}
\def\D{{\bf{D}}}

\begin{document}

\title{Features of electromagnetic waves in a complex plasma due to surface plasmon resonances on macroparticles}
\author{S. V. Vladimirov$^{1-3}$ \& O. Ishihara$^{3}$}

\affiliation{${}^{1}$ Metamaterials Laboratory, ITMO University, St. Petersburg, 199034, and Theoretical Department, Joint Institute of High Temperatures, Russian Academy of Sciences, Moscow, 125412, Russia}
\email{svvladi@gmail.com}

\affiliation{${}^{2}$School of Physics, The University of Sydney, NSW 2006, Australia}

\affiliation{${}^{3}$Yokohama National University, Yokohama, 240-8501, and Chubu University, Kasugai, 487-8501, Japan}
\email{oishihar@chubu.ac.jp}

\date{\today}

\begin{abstract}
The dielectric properties of complex plasma containing either metal or dielectric spherical inclusions (macroparticles, dust) are investigated. We focus on surface plasmon resonances on the macroparticle surfaces and their effect on electromagnetic wave propagation. It is demonstrated that the presence of surface plasmon oscillations significantly modifies plasma electromagnetic properties by resonances and cutoffs in the effective permittivity. This leads to related branches of electromagnetic waves and to the wave band gaps. The results are discussed in the context of dusty plasma experiments.
\end{abstract}
\maketitle

\pagebreak

\section{Introduction}

The presence of macroscopic particles (macroparticles, ''dust") significantly modifies plasma properties \cite{book1,book2,book3,ish1}, with visualization of kinetic effects of collective plasma dynamics \cite{ish2,ish3}. For example, the ''charging" damping appears due to plasma currents on dust particles \cite{tsythavn} thus affecting propagation and scattering of electromagnetic waves in such complex plasmas \cite{vl1,vl2}. Most of previous studies were focused on the effects of the presence of charged dust particles on plasma collective phenomena. Here, we demonstrate that another effect, namely, resonances with surface oscillations (surface plasmons) on macroparticle surfaces, can have important consequences on electromagnetic wave propagation in complex dusty plasmas.

Surface waves appearing on interfaces of two media with opposite signs of permittivities have long history of research \cite{boardman1,boardman2}. Surface plasma waves (sometimes called surface plasmons or surface plasmon polaritons), when one of the media is plasma, attracted special interest for many applications, for example, in plasmonics \cite{maradudin}, for light sources in such schemes as spasers \cite{sp_prl,sp_nature}, as well as for other applications such as nano-antennas, etc. \cite{prb}. The surface waves can transmit information on the properties of adjacent media, and therefore open a channel to influence electromagnetic properties in one of them by properties of another one \cite{sw_Physrep}.

In this paper, we investigate electromagnetic wave propagation in a plasma containing identical spherical macroparticles (dust). Two particular examples are considered in more details -- first, metallic dust, and second, dielectric dust particles. We simplify the problem by excluding effects of space dispersion (and therefore all related plasma phenomena such as Debye screening, sheath formation near the macroparticle surfaces, particle charging phenomena, etc.) but by keeping temporal dispersion. To calculate the effective permittivity of such a complex plasma, we employ the Maxwell Garnett (MG) approximation \cite{mg}, known for over a century and widely used in studies of two-component mixtures, considered as host medium and inclusions \cite{l2,l3,levy}.

The MG approximation involves exact calculation of the electromagnetic field induced in the host medium by a single spherical inclusion (in our case - spherical dust particle) and an approximate consideration of the electrostatic interaction between the inclusions (dust particles). In the lowest approximation (corresponding to the longest range interactions), the associate distortion of the electromagnetic field is caused by the charged dipoles induced on the other particles.

Our study demonstrates that due to surface plasmons on dust particles, new cutoffs and resonances appear in the electromagnetic wave propagation, as compared with the case when surface oscillations are ignored. This leads to new branches of electromagnetic waves in complex dusty plasma, not accounted before, and related band gaps. The presence of surface plasmon resonances makes complex plasma qualitatively different from a classical electron-ion plasma where no resonance are in the basic state (no external fields; resonances appear, e.g., in the presence of external magnetic field -- cyclotron resonance), and allows us to consider dusty plasma as a metamaterial (for electromagnetic waves with wavelengths much larger than distances between dust particles), either with near zero (ENZ) or very large epsilon (EVL) \cite{enz1,enz2,elv} (we note here that certain analogies also exist between magnetized plasmas and so-called wire media, actively studied in metamaterial applications \cite{belov,tysh}).

\section{The effective permittivity}

Consider a single sphere of radius $R$  filled by medium with permittivity $\veps_{in}=\veps_{in,0}-\omega_{in}^2/\omega^2$, i.e., we assume that in addition to permittivity $\veps_{in,0}={\rm const}\geq 1$ there is a number of free charges (''electrons") contributing to Drude-like dispersion with plasma frequency $\omega_{in}$ (and no losses). In two limiting cases we have: metal for $\veps_{in,0}=1$, $\omega_{in}\ne 0$ and dielectric for $\omega_{in}=0$ and $\veps_{in,0}=\veps_{id}={\rm const}\geq 1$ (vacuum when $\veps_{in,0}=1$). The sphere is embedded in a host medium with similar type of permittivity $\veps_h=\veps_{h,0}-\omega_h^2/\omega^2$, $\veps_{h,0}={\rm const}\geq 1$. Thus we account for two more cases: dielectric host with $\omega_h=0$ and $\veps_{h,0}=\veps_{hd}={\rm const}\geq 1$ (vacuum when $\veps_h=1$), and plasma with $\omega_h\ne 0$ and $\veps_{h,0}=1$. While taking into account the effects of temporal dispersion we ignore effects of space dispersion and other collective plasma phenomena (such as Debye screening, sheath formation, plasma charging currents onto sphere surface, etc.).

We assume that surface plasmon fields are electrostatic and oscillate with the frequency $\omega$ (i.e., their electrostatic field potential $\vphi\propto\sin(\omega t)$). In spherical geometry with azimuthal symmetry, solutions of the (linear) Poisson's equation ${\bf \nabla}\cdot\veps_{in,h}{\bf \nabla}\vphi=0$ are \cite{sw_Physrep}
\begin{eqnarray}
\vphi=\sum_{j=0}^{\infty}C_j(\omega)\left(r/R\right)^jP_j(\cs)\,, && \mbox{for $r<R$}\,,
\label{sol1}\\
\vphi=\sum_{j=0}^{\infty}C_j(\omega)\left(R/r\right)^{j+1}P_j(\cs)\,, && \mbox{for $r>R$}\,,
\label{sol2}
\end{eqnarray}
where $P_j$ are the Legendre polynomials and $C_j(\omega)$ are (arbitrary) functions of frequency $\omega$. The multipole expansion (\ref{sol1})--(\ref{sol2}) is in spherical harmonics and corresponds for $j=0$ (with $P_0=1$) to constant (on $r,\Theta$) potential inside the sphere, while for higher harmonics $j\geq 1$ -- to dipole field (with $P_1=\cs$) for $j=1$, quadrupole field for $j=2$, etc.

The boundary conditions are the continuity of potential $\vphi$ and normal (to the boundary, i.e., radial in the considered spherical geometry) component of displacement $\veps_{in,h}{\bf \nabla}\vphi$ across $r=R$. Thus we obtain dispersion equation for surface plasmons
\bq
j\veps_{in}=-(j+1)\veps_h\,.
\label{sweps}
\eq
For $j=0$ [const($r,\Theta$) potential inside the sphere], we have dispersion equation $\veps_h=0$ corresponding to bulk electrostatic plasma-type oscillations ($\omega^2=\omega_h^2/\veps_h$); the same for any $j$ and $\veps_{in,0}=\veps_{h,0}$, $\omega_h=\omega_{in}$ (no inclusion and therefore no boundary). For $j\geq 1$, solutions (surface plasmons) are possible only for opposite signs of $\veps_{in}$ and $\veps_h$ (i.e., there is no surface-type solutions when $\omega_{in}=\omega_h=0$ and both $\veps_{in,h,0}>0$). The (squared) eigenfrequencies of surface oscillations are
\bq
\omega_{s,j}^2=\frac{j\omega_{in}^2+(j+1)\omega_h^2}{j\veps_{in,0}+(j+1)\veps_{h,0}}\,.
\label{swfreq2}
\eq
In the limits of metal sphere in vacuum ($\veps_{in,0}=\veps_{h,0}=1$, $\omega_h=0$, $\omega_{in}\ne 0$) and vacuum spherical void in plasma ($\veps_{in,0}=\veps_{h,0}=1$, $\omega_{in}=0$, $\omega_h\ne 0$), respectively, we have known \cite{sw_Physrep} expressions for the eigenfrequencies of surface plasmon dipole ($j=1$) oscillations $\omega_{sp,in}=\omega_{in}/\sqrt{3}$ and $\omega_{sp,h}=\omega_h\sqrt{2/3}$. The factors $1/\sqrt{3}$ or $\sqrt{2/3}$ are different from $1/\sqrt{2}$ for surface oscillations on the boundary of semi-infinite plasma \cite{sw_Physrep}. To proceed correctly to the limit of infinitely large radius, we need to sum up all (the infinite number of) spherical harmonics contributing to the resulting factor $1/\sqrt{2}$ for the semi-infinite plasma (like calculating plasma permittivity in infinitely small magnetic field, where all cyclotron harmonics have to be sum up in order to obtain the correct field-free limit \cite{abr}).

In the presence of external (with respect to the sphere) electromagnetic wave field, we follow Garnett \cite{mg} and Landau \cite{llecm} and proceed only in the dipole approximation (i.e., for $j=1$ in the above example of electrostatic fields) which corresponds to the longest possible range of surface fields. For the electric field $\E_{in}$ and the displacement $\D_{in}$ inside a single sphere in the presence of external field $\E_0$, we have \cite{llecm}
\bq
\D_{in}+2\veps_h\E_{in}=3\veps_h\E_0\,.
\label{displ}
\eq
This relation allows us to calculate the induced dipole moment of the sphere and obtain, for a number of identical non-interacting spheres, the so-called dilute limit for the effective permittivity \cite{mg}
$\tilde{\veps}_h^{\rm dil}=\veps_h[1+3f(\veps_{in}-\veps_h)/(\veps_{in}+2\veps_h)]$, where $f=V_{in}/V_h$, $V=V_{in}+V_h$, is the volume fraction occupied by spherical inclusions (e.g., dust in plasma). Obviously, the volume fraction should be sufficiently small here in order to neglect interactions between spheres.

To account for dipolar interactions between spheres, we apply the excluded volume approach \cite{brpipp} where the field acting on inclusion particles (e.g., dust particles) is the average host field $\E_h$, which is now appearing on the r.h.s. of Eq. (\ref{displ}) instead of $\E_0$. The field $\E_h$ differs from $\E_0$ due to the correlations between positions of different (non-overlapping) spheres. The averaged field over the entire system, inside as well as outside the spherical inclusions (dust particles), is $\E_0$. Thus $f\E_{in}+(1-f)\E_h=\E_0$ and for $\E_h$ we now have $\E_h=\E_0/(1-f+3f\veps_h/(\veps_{in}+2\veps_h))$. Calculating the volume (average) polarization, we finally arrive at MG result \cite{mg}
\bq
\tilde{\veps}_h=
\veps_h\left[1+\frac{3f(\veps_{in}-\veps_h)/(\veps_{in}+2\veps_h)}{1-f+3f\veps_h/(\veps_{in}+2\veps_h)}\right]
\equiv\veps_h\frac{1+2\tilde{f}}{1-\tilde{f}}\,,\;\;\;\;\;\;\;\;
\tilde{f}=f\frac{\veps_{in}-\veps_h}{\veps_{in}+2\veps_h}\,,
\label{epseff}
\eq
where we defined the ``impedance" volume fraction $\tilde{f}$. The resonance in the effective permittivity corresponds to $1=\tilde{f}$ while the cutoff -- to $1=-2\tilde{f}$. We note that there is a number of generalizations of MG result (\ref{epseff}), e.g., for anisotropic material of inclusions \cite{levy}. For our purposes here, however, it is sufficient to assume the material isotropy of inclusions (dust particles), as well as isotropic character of embedding host medium (e.g., gaseous plasma in dusty plasma experiments).

In the limit $f\to 0$ (no inclusions in the medium, such as no dust in plasma) we have from (\ref{epseff}) $\tilde{\veps}_h\to\veps_h$. In the opposite (formal) limit $f\to 1$ (inclusions occupy all space in the host medium) we have from (\ref{epseff}) $\tilde{\veps}_h\to\veps_{in}$, respectively.

\section{The dispersion equation}

Knowing the effective permittivity, we apply the dispersion equation for electromagnetic waves $\tilde{\veps}_h=k^2c^2/\omega^2\equiv n^2$, where $n$ stands for the refraction index. This dispersion equation can be written as (we remind that damping is ignored here)
\bq
\left(1-\frac{\tilde{\omega}_h^2}{\omega^2}\right)\left(1-\frac{\omega_{\rm cut}^2}{\omega^2}\right)=
\tilde{n}^2\left(1-\frac{\omega_{\rm res}^2}{\omega^2}\right)\,.
\label{dispeq}
\eq
Here, $\tilde{\omega}_h^2=\omega_h^2/\veps_{h,0}$ and $\tilde{n}^2=k^2\tilde{c}^2/\omega^2$, with decreased, due to the presence of dielectrics $\veps_{h,0}$ and $\veps_{in,0}$, speed of light: $\tilde{c}^2=c^2/\tilde{\veps}_{h,0}$, where (compare with expression (\ref{epseff}))
\bq
\tilde{\veps}_{h,0}=\veps_{h,0}\frac{1+2\tilde{f}_0}{1-\tilde{f}_0}\,,\;\;\;\;\;\;\;\;
\tilde{f}_0=f\frac{\veps_{in,0}-\veps_{h,0}}{\veps_{in,0}+2\veps_{h,0}}\,.
\label{tildeeps0}
\eq
When $f\to 0$ (no inclusions in the medium), we have $\tilde{\veps}_{h,0}\to\veps_{h,0}$; in the opposite (formal) limit $f\to 1$ (when inclusions occupy all space and there is no host medium) we have expected result $\tilde{\veps}_{h,0}\to\veps_{in,0}$ (note that the dispersion equation is also changed in these two limiting cases, see below).

In Eq. (\ref{dispeq}), we have defined two new characteristic frequencies related to the presence of surface plasmon oscillations on inclusion particle surfaces: the cutoff frequency $\omega_{\rm cut}$ and the resonance frequency $\omega_{\rm res}$. We have
\bq
\omega_{\rm cut}^2=\omega_{sp}^2+
\frac{6f\veps_{h,0}(\omega_{sp}^2-\tilde{\omega}_h^2)}{\veps_{in,0}+2\veps_{h,0}+2f(\veps_{in,0}-\veps_{h,0})}\,,
\label{cut}
\eq
where the eigenfrequency of (dipolar) surface plasmons $\omega_{sp}^2=\omega_{s,1}^2$, see (\ref{swfreq2}). Furthermore,
\bq
\omega_{\rm res}^2=\omega_{sp}^2-
\frac{3f\veps_{h,0}(\omega_{sp}^2-\tilde{\omega}_h^2)}{\veps_{in,0}+2\veps_{h,0}-f(\veps_{in,0}-\veps_{h,0})}\,.
\label{res}
\eq
For $f\to 0$, from (\ref{cut}) and (\ref{res}) we have $\omega_{\rm cut}\to\omega_{\rm res}\to\omega_{sp}$, this corresponds to the transformation of dispersion equation (\ref{dispeq}) to the inclusion-free form: $\veps_h=n^2$. For $f\to 1$, we have from (\ref{cut}) $\omega_{\rm cut}\to\tilde{\omega}_{in}\equiv\omega_{in}/\sqrt{\veps_{in,0}}$ and from (\ref{res}) $\omega_{\rm res}\to\tilde{\omega}_h$, this corresponds to the transformation of dispersion equation (\ref{dispeq}) to the pure inclusion-only form: $\veps_{in}=n^2$.

Consider three limiting cases: (a), metal spherical inclusions in gaseous plasma, $\veps_{in,0}=\veps_{h,0}=\tilde{\veps}_{h,0}=1$, $\tilde{\omega}_{in}=\omega_{pm}$, $\tilde{\omega}_h=\omega_{pe}$, $\omega_{sp}^2=(\omega_{pm}^2+2\omega_{pe}^2)/3$, where $\omega_{pm(e)}$ is the plasma frequency of metal (plasma) electrons; (b), metal spherical inclusions in dielectric medium, $\veps_{in,0}=1$, $\veps_{h,0}=\veps_{hd}=$const, $\tilde{\omega}_{in}=\omega_{pm}$, $\tilde{\omega}_h=0$, $\omega_{sp}^2=\omega_{pm}^2/(1+2\veps_{hd})$; and (c), dielectric spherical inclusions in plasma, $\veps_{in,0}=\veps_{id}=$const, $\veps_{h,0}=1$, $\tilde{\omega}_{in}=0$, $\tilde{\omega}_h=\omega_{pe}$, $\omega_{sp}^2=2\omega_{pe}^2/(\veps_{id}+2)$.

In the case (a), we have
\bq
\omega_{\rm cut}^2=\omega_{sp}^2+2f\left(\omega_{sp}^2-\omega_{pe}^2\right)\,,\,\,\,\,
\omega_{\rm res}^2=\omega_{sp}^2-f\left(\omega_{sp}^2-\omega_{pe}^2\right)\,.
\label{metpl}
\eq
In the case (b), we obtain
\bq
\omega_{\rm cut}^2=\omega_{sp}^2\left[1+\frac{6f\veps_{hd}}{1+2\veps_{hd}+2f(1-\veps_{hd})}\right]\,,\,\,\,\,
\omega_{\rm res}^2=\omega_{sp}^2\left[1-\frac{3f\veps_{hd}}{1+2\veps_{hd}-f(1-\veps_{hd})}\right]\,.
\label{metdiel}
\eq
For $\veps_{hd}=1$, these expressions become $\omega_{\rm cut}^2=\omega_{sp}^2(1+2f)$ and $\omega_{\rm res}^2=\omega_{sp}^2(1-f)$, cf. (\ref{metpl}) with $\omega_{pe}=0$. Finally, in the case (c) we have
\bq
\omega_{\rm cut}^2=\omega_{sp}^2\left[1-\frac{3f\veps_{id}}{\veps_{id}+2+2f(\veps_{id}-1)}\right]\,,\,\,\,\,
\omega_{\rm res}^2=\omega_{sp}^2\left[1+\frac{3f\veps_{id}/2}{\veps_{id}+2-f(\veps_{id}-1)}\right]\,.
\label{dielpl}
\eq
For $\veps_{id}=1$, these expressions become $\omega_{\rm cut}^2=\omega_{sp}^2(1-f)$ and $\omega_{\rm res}^2=\omega_{sp}^2(1+f/2)$.

It is important for the wave dispersion that in general for $\tilde{\omega}_{in}>\tilde{\omega}_h$, the cutoff frequency is above the resonance frequency; moreover, we have $\tilde{\omega}_{in}>\omega_{\rm cut}>\omega_{sp}>\omega_{\rm res}>\tilde{\omega}_h$. In the opposite case $\tilde{\omega}_{in}<\tilde{\omega}_h$, there is in general $\tilde{\omega}_{in}<\omega_{\rm cut}<\omega_{sp}<\omega_{\rm res}<\tilde{\omega}_h$. We can clearly see that in the above particular examples (a)--(c).

\section{The wave dispersion}

Consider $0<f<1$, i.e., exclude degenerate cases $f=0$ and $f=1$. Dispersion equation (\ref{dispeq}) has then a typical diagram of cutoffs and resonances, see Figs. 1 and 2, respectively, for the cases $\tilde{\omega}_{in}>\tilde{\omega}_h$ and $\tilde{\omega}_{in}<\tilde{\omega}_h$ (such as metallic and dielectric spheres (dust) immersed in gaseous plasma). Solutions of dispersion equation (\ref{dispeq}) correspond to two branches (real frequency, $\omega^2\geq 0$, and refraction index, $\tilde{n}^2\geq 0$) for both cases.

\begin{figure}[h]
\includegraphics[width=8 cm]{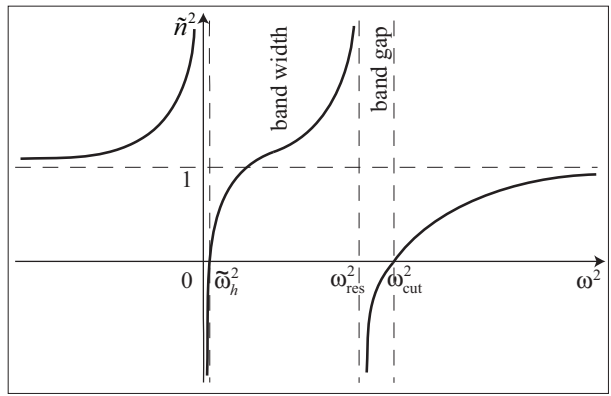}
\caption{Qualitative behavior of the squared index of refraction $\tilde{n}^2$  vs squared frequency $\omega^2$ for electromagnetic wave in the case $\tilde{\omega}_{in}>\tilde{\omega}_h$. The lower cutoff is at $\tilde{\omega}_h$.}
\label{fig0_1}
\end{figure}
For $\tilde{\omega}_{in}>\tilde{\omega}_h$, the dependence of $\tilde{n}^2$ on $\omega^2$, see Fig. 1, shows the first, low frequency, branch starting from the host medium plasma frequency $\tilde{\omega}_{h}^2$. This branch begins as the standard plasma branch in the long-wavelength limit, however, then asymptotically approaches the (squared) resonance frequency $\omega_{\rm res}^2$ (\ref{res}) at which $\tilde{n}^2\to+\infty$, and this frequency is associated with the surface plasmon resonances. The second, high frequency branch, in the long-wavelength limit associated with surface plasmons, starts at the (squared) cutoff frequency $\omega_{\rm cut}^2$ (\ref{cut}), and then approaches the light line $\tilde{n}^2=1$; in the short-wavelength limit this branch loses its association with surface plasmons and becomes standard vacuum light line. Thus the presence of surface plasmon oscillations changes the plasma branch by interrupting it at the surface plasmon resonance; as a result two branches appear asymptotically approaching the electromagnetic plasma wave in the low-frequency and the high-frequency limits, respectively. The band gap is defined by $\Delta\omega_{g,1}=\omega_{\rm cut}-\omega_{\rm res}$ which for $f\ll 1$ is small, $\Delta\omega_{g,1}\sim f\omega_{sp}\ll\omega_{sp}$. The finite band width of the first branch (the second branch has infinite band width) is defined by $\Delta\omega_{w,1}=\omega_{\rm res}-\tilde{\omega}_h$. For $f\ll 1$, while the band gap is narrow $\sim f\omega_{sp}$, the band width is finite of order $\omega_{sp}$.

\begin{figure}[t]
\includegraphics[width=8 cm]{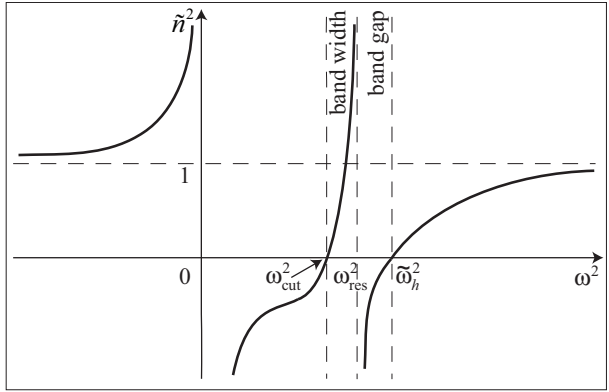}
\caption{Qualitative behavior of the squared index of refraction $\tilde{n}^2$ vs squared frequency $\omega^2$ for electromagnetic wave in the case $\tilde{\omega}_{in}<\tilde{\omega}_h$.}
\label{fig0_2}
\end{figure}
For $\tilde{\omega}_{in}<\tilde{\omega}_h$, the dependence of $\tilde{n}^2$ on $\omega^2$ is shown in Fig. 2. The first, low-frequency, branch is now completely related to surface plasmons; it starts from the (squared) cutoff frequency $\omega_{\rm cut}^2$ (\ref{cut}); then this branch asymptotically approaches the (squared) resonance frequency $\omega_{\rm res}^2$ (\ref{res}) at which $\tilde{n}^2\to+\infty$. The second, high frequency branch is the standard host (e.g., gaseous plasma) branch in this case; it starts at the (squared) frequency $\tilde{\omega}_h^2$ and then approaches the light line $\tilde{n}^2=1$ as the (squared) frequency tends to infinity. Thus we see that in this case the presence of surface plasmon resonances does not affect the existence of host (plasma) branch but adds completely new branch fully related to surface plasmon oscillations. The band gap is defined by $\Delta\omega_{g,2}=\tilde{\omega}_h-\omega_{\rm res}$. The band width of the first branch associated with surface plasmons (the second plasma branch has infinite band width) is defined by $\Delta\omega_{w,2}=\omega_{\rm res}-\omega_{\rm cut}$. For $f\ll 1$, the band width is becoming negligibly narrow $\Delta\omega_{w,2}\sim f\omega_{sp}\ll\omega_{sp}$ while the band gap is finite $\Delta\omega_{g,2}\sim\omega_{sp}$.

For dispersion curves, we have in the limit of small wave numbers for both cases $\tilde{\omega}_{in}>(<)\tilde{\omega}_h$, Figs. 1 and 2, the following approximate expressions for the surface plasmon (sp) related and host (plasma) related branches:
\bq
\omega^2\Biggr|_{kc\to 0} \approx\left\{
\begin{array}{ll}
\omega_{\rm cut}^2+k^2\tilde{c}^2(\omega_{\rm cut}^2-\omega_{\rm res}^2)/(\omega_{\rm cut}^2-\tilde{\omega}_h^2)\,, & \mbox{sp branch},\\
\tilde{\omega}_h^2+k^2\tilde{c}^2(\omega_{\rm res}^2-\tilde{\omega}_h^2)/(\omega_{\rm cut}^2-\tilde{\omega}_h^2)\,, & \mbox{host (plasma) branch}.
\end{array}
\right.
\label{smallk}
\eq
In in the limit of large wave numbers, we obtain
\bq
\omega^2\Biggr|_{kc\to \infty} \approx\left\{
\begin{array}{ll}
k^2\tilde{c}^2+\tilde{\omega}_h^2+\omega_{\rm cut}^2-\omega_{\rm res}^2\,, & \mbox{em branch},\\
\omega_{\rm res}^2-(\omega_{\rm cut}^2-\tilde{\omega}_h^2)(\omega_{\rm cut}^2-\omega_{\rm res}^2)/k^2\tilde{c}^2\,, & \mbox{sp resonance branch}.
\end{array}
\right.
\label{largek}
\eq
Remind that for $\tilde{\omega}_{in}>\tilde{\omega}_h$ and $kc\to\infty$, the high frequency (hf) sp branch is converted to hf electromagnetic (em) branch and the low frequency (lf) host (plasma) branch is converted to sp resonance branch.

For the limiting case (a), metal spherical inclusions in gaseous plasma, $\veps_{in,0}=\veps_{h,0}=\tilde{\veps}_{h,0}=1$, $\tilde{\omega}_{in}=\omega_{pm}$, $\tilde{\omega}_h=\omega_{pe}$, $\omega_{sp}^2=(\omega_{pm}^2+2\omega_{pe}^2)/3$, $\tilde{c}=c$, we calculated the refraction index, see Fig. 3.
\begin{figure}[t]
\includegraphics[width=7 cm]{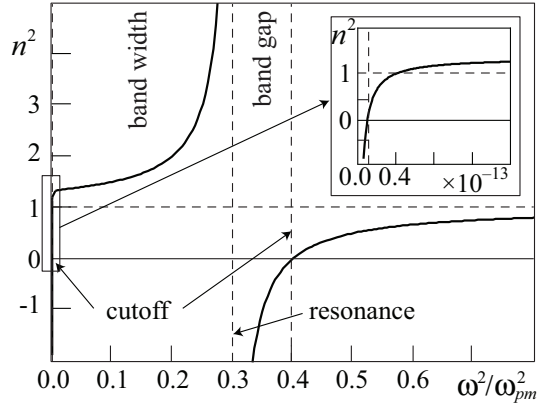}
\caption{Calculated squared index of refraction $n^2$ vs squared frequency $\omega^2$ of electromagnetic wave in a plasma with metallic spherical inclusions (metallic dust). The frequency is normalized by the plasma frequency of electrons in metal $\omega_{pm}$; $\omega_{pe}/\omega_{pm}=10^{-7}$, $f=0.1$. The inset shows the behavior in the low frequency and low wave number domain.}
\label{fig1}
\end{figure}
The calculated dispersion curves are shown in Fig. 4.
\begin{figure}[b]
\includegraphics[width=7.5 cm]{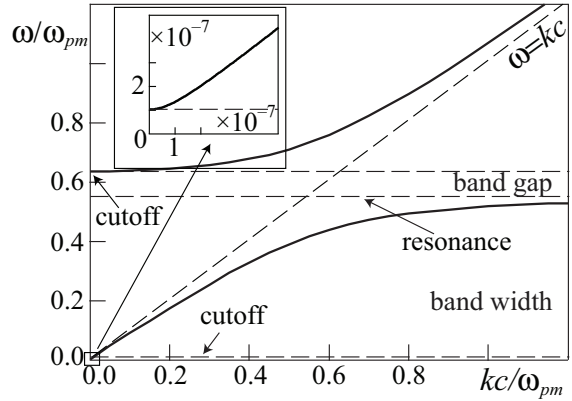}
\caption{Calculated dispersion relation of electromagnetic wave in a plasma with spherical metallic inclusions (metallic dust). The frequency is normalized by the plasma frequency of electrons in metal $\omega_{pm}$, the wave number is normalized by $\omega_{pm}/c$; $\omega_{pe}/\omega_{pm}=10^{-7}$, $f=0.1$. The inset shows the behavior in the low frequency and low wave number domain.}
\label{fig3}
\end{figure}
In this case, we have from (\ref{smallk}) and (\ref{largek})
\bq
\omega^2\Biggr|_{kc\to 0} \approx\left\{
\begin{array}{ll}
\omega_{sp}^2+2f\left(\omega_{sp}^2-\omega_{pe}^2\right)+3fk^2c^2/(1+2f)\,, & \\
\omega_{pe}^2+k^2c^2(1-f)/(1+2f)\,, &
\end{array}
\right.
\label{smallkmp}
\eq
\bq
\omega^2\Biggr|_{kc\to \infty} \approx\left\{
\begin{array}{ll}
k^2c^2+f\omega_{pm}^2+(1-f)\omega_{pe}^2\,, & \\
\omega_{sp}^2-f(\omega_{sp}^2-\omega_{pe}^2)-3f(1+2f)(\omega_{sp}^2-\omega_{pe}^2)^2/k^2c^2\,. &
\end{array}
\right.
\label{largekmp}
\eq
Note that expression $f\omega_{pm}^2+(1-f)\omega_{pe}^2$ has simple physical meaning -- it is related to the total number of electrons, from metal $N_m=V_mn_{e,m}=Vfn_{e,m}$ and plasma $N_p=V_pn_{e,p}=V(1-f)n_{e,p}$, contributing to the corresponding plasma frequency, $\omega_{pe,m}^2=4\pi n_{e,p,m}e^2/m_e$. Here, $n_{e,p,m}$ is the respective number density of plasma $p$ or metal $m$ electrons with charge $e$ and mass $m_e$.

For the limiting case (b), metal spherical inclusions in dielectric medium, $\veps_{in,0}=1$, $\veps_{h,0}=\veps_{hd}=$const, $\tilde{\omega}_{in}=\omega_{pm}$, $\tilde{\omega}_h=0$, $\omega_{sp}^2=\omega_{pm}^2/(1+2\veps_{hd})$, $\tilde{c}^2=c^2/\tilde{\veps}_{hd}$, $\tilde{\veps}_{hd}=\veps_{hd}(1+2\veps_{hd}+2f(1-\veps_{hd}))/(1+2\veps_{hd}-f(1-\veps_{hd}))$ we obtain (note that for host em branch we have the dielectric light line) from (\ref{smallk}) and (\ref{largek})
\bq
\omega^2\Biggr|_{kc\to 0} \approx\left\{
\begin{array}{ll}
\omega_{sp}^2+3f
[2\veps_{hd}\omega_{sp}^2+3k^2c^2]/(1+2\veps_{hd}+2f(1-\veps_{hd}))\,, & \\
k^2c^2(1-f)/\veps_{hd}(1+2f)\,, &
\end{array}
\right.
\label{smallkmd}
\eq
\bq
\omega^2\Biggr|_{kc\to\infty} \approx\left\{
\begin{array}{ll}
k^2\tilde{c}^2+9f\veps_{hd}\omega_{sp}^2/(1+2\veps_{hd})
(1+2\tilde{f}_{hd})(1-\tilde{f}_{hd})\,, & \\
\omega_{sp}^2\left[1-f
-9f\veps_{hd}(1+2f)\omega_{sp}^2/k^2\tilde{c}^2(1+2\tilde{f}_{hd})^2(1+2\veps_{hd})\right]/(1-\tilde{f}_{hd})\,, &
\end{array}
\right.
\label{largekmd}
\eq
where from Eq. (\ref{tildeeps0}) we have $\tilde{f}_{hd}=f(1-\veps_{hd})/(1+2\veps_{hd})$ in the considered case. Note that for metal spheres in vacuum $\veps_{hd}=1$, Eqs. (\ref{smallkmd})--(\ref{largekmd}) coincide with Eqs. (\ref{smallkmp})--(\ref{largekmp}) where $\omega_{pe}=0$.

\begin{figure}[h]
\includegraphics[width=7 cm]{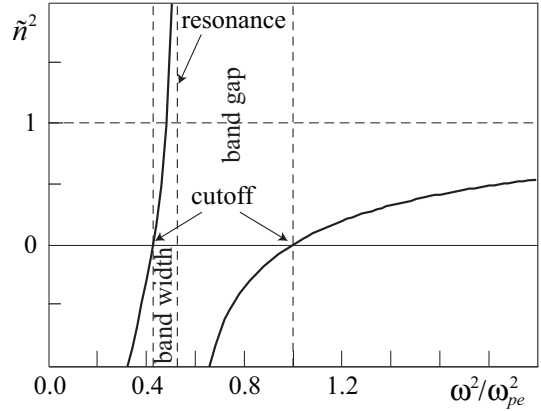}
\caption{Calculated squared index of refraction $\tilde{n}^2$ vs squared frequency $\omega^2$ of electromagnetic wave in a plasma with dielectric spherical inclusions (dielectric dust). The frequency is normalized by the plasma frequency $\omega_{pe}$; $\veps_{id}=2$, $f=0.1$.}
\label{fig2}
\end{figure}
Finally, for the limiting case (c), dielectric spherical inclusions (dust) in plasma, $\veps_{in,0}=\veps_{id}=$const, $\veps_{h,0}=1$, $\tilde{\omega}_{in}=0$, $\tilde{\omega}_h=\omega_{pe}$, $\omega_{sp}^2=2\omega_{pe}^2/(\veps_{id}+2)$, $\tilde{c}^2=c^2/\tilde{\veps}_{id}$, $\tilde{\veps}_{id}=(\veps_{id}+2+2f(\veps_{id}-1))/(\veps_{id}+2-f(\veps_{id}-1))$, we calculated the refraction index, see Fig. 5. The calculated dispersion curves are shown in Fig. 6.
\begin{figure}[h]
\includegraphics[width=7.5 cm]{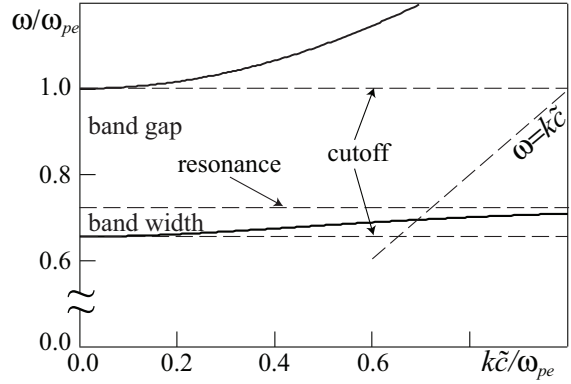}
\caption{Calculated dispersion relation of electromagnetic wave in a plasma with dielectric spherical inclusions (dielectric dust). The frequency is normalized by the electron plasma frequency $\omega_{pe}$, the wave number is normalized by $\omega_{pe}/\tilde{c}$, with normalized speed of light $\tilde{c}^2=c^2/\tilde{\veps}_{id}$, $\tilde{\veps}_{id}=(\veps_{id}+2+2f(\veps_{id}-1))/(\veps_{id}+2-f(\veps_{id}-1))$; $\veps_{id}=2$, $f=0.1$.}
\label{fig4}
\end{figure}
In this case, we have
\bq
\omega^2\Biggr|_{kc\to 0} \approx\left\{
\begin{array}{ll}
\omega_{pe}^2+k^2c^2(1-f)/(1+2f)\,, & \\
\omega_{sp}^2\left[1-f+9fk^2c^2/2\omega_{pe}^2(1+2f)\right]/(1+2\tilde{f}_{id})\,, &
\end{array}
\right.
\label{smallkdp}
\eq
\bq
\omega^2\Biggr|_{kc\to\infty} \approx\left\{
\begin{array}{ll}
k^2\tilde{c}^2+\omega_{pe}^2+9f\veps_{id}\omega_{sp}^2/2(\veps_{id}+2)
(1+2\tilde{f}_{id})(1-\tilde{f}_{id})\,, & \\
\omega_{sp}^2\left[1+(f/2)
-9f\veps_{id}^2\omega_{sp}^2(1+2f)/2k^2\tilde{c}^2(\veps_{id}+2)(1+2\tilde{f}_{id})\right]/(1-\tilde{f}_{id})\,, &
\end{array}
\right.
\label{largekdp}
\eq
where from Eq. (\ref{tildeeps0}) we have $\tilde{f}_{id}=f(\veps_{id}-1)/(\veps_{id}+2)$ in the considered case.

\section{Discussion and conclusion}

Above, we have neglected potential losses in the considered media. While wave damping in dielectrics and gaseous plasma can be considered negligible, losses in metal are higher. In metals, the electron number density is normally within the range $n_{em}\sim1\div 10\times 10^{22}$ cm$^{-3}$ \cite{ashcroft}. For the Drude approach, the plasma frequency is then $f_{pm}=\omega_{pm}/2\pi\approx 9000\sqrt{n_{em}}\sim 10^{16}$Hz. On the other hand, the relaxation times are typically $\tau_m\sim 1\div 10\times 10^{-15}$s \cite{ashcroft}; this gives for $\omega_{pm}\tau_m\sim 1\div 10\times 10^2$. In this regard, the important parameter is $\Delta\omega_{g,m}\tau_m$; note that the finite band width of the lf branch is of order $\omega_{sp}\sim\omega_{pm}$ (in experiments with metal particles in gaseous plasma, we have $\omega_{pm}\gg\omega_{pe}$ since $f_{pe}$ is typically of order few GHz)) and therefore its relation to $\tau_m$ is approximately the same as for $\omega_{pm}\tau_m$. We have $\Delta\omega_{g,m}=\omega_{\rm cut}-\omega_{\rm res}\approx f\omega_{pm}$, see (\ref{metpl}), and therefore $\Delta\omega_{g,m}\tau_m\approx f\omega_{pm}\tau_m$. Since in plasma experiments the volume fraction of dust particles can be very small, $f\lesssim 10^{-2}$, then we have $f\omega_{pm}\tau_m\lesssim 1$. This means that the band gap and thus the whole surface plasmon resonance may be completely smeared off due to high damping on dust metal particle surfaces, and as a result we have standard plasma em branch (as without dust) with additional damping due to damping of surface plasmons. On the other hand, for the case (b) of metal particles in dielectric, the restriction due to the surface plasmon damping is not such strong for possible larger volume fractions $f$.

The considered approach is for relatively large wavelengths of em waves, $\lambda_w\gg d$ where $d$ is the distance between inclusions (dust particles). Moreover, the wave lengths should exceed typical effects of space dispersion. For gaseous plasma and typical dusty plasma experiments, Debye length $\lambda_{De}$ stands approximately for both, we have $\lambda_{De}\sim 10^{-2}$cm. Thus of most interest in the considered model for dusty plasma experiments are not the optical em wave frequencies with $\lambda_w\sim 0.5\times 10^{-4}$cm, but much lower frequencies, with $\lambda_w\sim 10^{-1}$cm. This range approximately corresponds to the new (due to surface plasmon resonances) wave band in the case of dielectric particles in plasma, see the above case (c). Note that this wavelengths are still much less than the typical dust cloud size of order 10 cm.

In dusty plasmas, the dust inclusions are charged due to charging by various processes; in laboratory plasmas, this is mostly by plasma currents \cite{book1,book2,book3}. The presence of dust charges affects the boundary condition for surface wave fields. However, the most effective for the considered processes would have been very high frequency dust charge oscillations, in the range of the surface plasmon frequency $\omega_{sp}$. Even in the case of the lowest possible $\omega_{sp}$, the considered case (c) of dielectric spherical inclusions in gaseous plasma, $\omega_{sp}\sim \omega_{pe}$, i.e., much higher than the typical frequency of dust charge variations $\nu_{ch}$ which is affected by ion processes as well \cite{tsythavn} and is entirely in completely different frequency range: $\nu_{ch}\ll\omega_{pe}$. Thus the charges can be considered as constant in our approximation, and their influence on the high frequency wave processes is indirect. Indeed, the charges (as well as other collective plasma processes such as formation of plasma sheath near particle surfaces) affect the interparticle separation and therefore the volume fraction occupied by dust.

To conclude, the presence of inclusions in a medium is found to produce cutoffs and resonance in the propagation of electromagnetic waves. The surface plasmon oscillation on the surfaces of the inclusions (macroparticles, dust) in the medium (plasma) is responsible for the splitting of the electromagnetic branch and the emergence of two branches in the propagation of electromagnetic waves in the medium. The high frequency branch and the low frequency branch are separated by the band gap on the order of the frequency of the surface plasmon oscillations; the band width for the lower frequency branch is finite while the upper frequency band has infinite band width. A complex plasma with metallic dust in a plasma is characterized by the resonance frequency of the order of the surface plasmon frequency of metals. On the other hand, complex plasma with dielectric dust is characterized by the resonant frequency of the order of the plasma frequency of electrons in the plasma. The band gaps are controlled by the number density of inclusions in plasma.


%
\acknowledgments
This study was partially supported by the Ministry of Education and Science of the Russian Federation, and by the Asian Office of Aerospace R $\&$ D under Grant No. FA2386-14-1-4021.


%

\end{document}